\documentstyle[12pt]{article}
\def\be{\begin{equation}}
\def\ee{\end{equation}}
\def\a{\alpha}
\def\b{\beta}
\def\e{\varepsilon}

\def\La{\Lambda}

\def\O{\Omega}
\def\ro{\varrho}
\begin{document}
 \renewcommand{\theequation}{\thesection.\arabic{equation}} 
 \title{The tetralogy of Birkhoff  theorems} 
 \author{Hans-J\"urgen  Schmidt} %
\date{October 21, 2012}
\maketitle
\centerline{Institut f\"ur Mathematik, Universit\"at Potsdam, Germany} 
\centerline{Am Neuen Palais 10, D-14469 Potsdam,  \  hjschmi@rz.uni-potsdam.de}

\begin{abstract}
We classify the existent Birkhoff-type theorems into four classes: First, in field theory, 
the theorem states the absence of helicity 0- and spin 0-parts of the gravitational field. 
 Second, in relativistic astrophysics, it is the statement that the gravitational far-field 
 of a spherically symmetric star carries, apart from its  mass, no information about the star; 
 therefore,  a radially oscillating star has a static gravitational far-field. Third, in 
mathematical physics, Birkhoff's theorem reads: up to singular exceptions of measure
 zero, the spherically symmetric solutions of Einstein's vacuum field equation with 
 $\La=0$ can be expressed by the Schwarzschild metric; for  $\La \ne  0$, it is the
 Schwarzschild-de Sitter metric instead. Fourth, in differential geometry, any statement 
of the type: every member of a family of pseudo-Riemannian space-times 
has more isometries than expected from the original metric ansatz, carries the 
name Birkhoff-type theorem. Within the fourth of these classes we present some new 
results with further values of dimension and signature of the related spaces; including
 them are some counterexamples: families of space-times where no Birkhoff-type 
theorem  is valid. These counterexamples further confirm the conjecture, that the 
Birkhoff-type theorems have their origin in the property, that the two eigenvalues
of the Ricci tensor  of two-dimensional pseudo-Riemannian spaces always coincide,
a property not having an analogy in higher dimensions. Hence, Birkhoff-type theorems
exist only for those physical situations which are  reducible to two dimensions. 
\end{abstract}

\vspace{5.mm} Keyword(s): Birkhoff  theorem, Einstein space, isometry group 

\section{Introduction}
\setcounter{equation}{0} 

Four different types of theorems carry the name Birkhoff  theorem, all of them  refer 
to the original Birkhoff result from 1923, see  \cite{b1} for a presentation  of  the earlier 
papers about it. All of them are in a sense related to the spherically symmetric metric
 \be 
ds^2 =  - \left(1 - \frac{2m}{r}\right)dt^2   +  \frac{dr^2}{1 - 2m/r} + r^2 
\left( d\psi^2 + \sin^2 \psi d\varphi^2 \right)
\ee
 and its property  that the $t$-translation  $\partial/\partial t$ represents an isometry.
 It  is one of the possibilities to write  the metric of the Schwarzschild black hole
and is valid for all points of the space-time except the horizon at $r =2m$. 
 \par
First, in field theory, the theorem states the absence of spin 0-parts of the 
gravitational field within Einstein's theory following from the Einstein-Hilbert 
lagrangian $L_{\rm EH} = R$. One of its possible  counterparts is a theory, where
the lagrangian $L_{\rm FOG} = L_{\rm EH} +  l^2 R^2$ has a spin 0-part stemming from
the $R^2$-term. These considerations may be restricted to the linearized field equations,
 where closed-form solutions are available.  Accordingly, even in the linearized solutions, 
   $\partial/\partial t$ fails to be  an isometry  in fourth-order gravity defined by 
 $L_{\rm  FOG}$. The short-hand description of Birkhoff's theorem with the words:
 spherically symmetric solutions of  Einstein's field equation  are static is a little bit
 dangerous, as it may lead to misunderstandings  about the validity  of metric (1.1) in
 the region $0 < r < 2m$, where $t$ fails to be a timelike coordinate. The best 
 version to circumvent this misunderstanding is to say, that spherically symmetric 
 solutions of  Einstein's field equation possess a fourth isometry, represented by the
hypersurface-orthogonal Killing vector $\partial/\partial t$. However, this formulation  is 
not very convenient, so another  possibility  has been used  in some places: There a 
 space-time is defined to be spherically symmetric, if  it has  a  SO(3)-group of isometries, 
whose orbits are isomorphic to a standard two-sphere $S^2$ with area  $4 \pi r^2$, and 
the gradient of the scalar  $r$ represents  a spacelike vector vanishing at the center of 
symmetry only.  With this definition of spherical symmetry, the original Birkhoff 
formulation  keeps valid. Another possibility to circumvent the region with  $ r \le 2m$
 applied in field theory is to concentrate on the linearized field equation with 
smallness parameter being essentially $m$, there a Fourier  transform of the solutions  is  
 possible.  \par
Second, in relativistic astrophysics, the main emphasis is on the formulation: the set of 
spherically symmetric vacuum solutions of   Einstein's field equation can be 
parametrized by one single parameter $m$. Of course, for $m =0$, metric (1.1) 
 represents nothing but the flat Minkowski space-time of special relativity theory. 
  Sometimes one can see formulations like: In all other cases, $m>0$ represents
 the total mass of the central object. Again, such a formulation, though justified, is a 
little bit dangerous, as also for $m<0$, metric (1.1) represents a vacuum solution. 
 Formally, this case can be subsumed under the $m>0$-case, if one allows also 
negative values of $r$ in that metric. But the real astrophysical reason for  restricting 
$m$ to values $m \ge 0$ is the fact, that  all objects  composed of  normal matter have a 
positive  total mass. The importance of the Birkhoff theorem  is the following: the 
gravitational  far-field of a spherically symmetric star carries, apart from its total mass, 
absolutely no information about the structure of   the star, so e.g. a radially oscillating star 
has a static  gravitational far-field.  This property was already known to be valid in 
Newton's theory of gravitation, but it came as a surprise that such a property will also be 
valid in Einstein's theory, where  the gravitational field, the metric $g_{AB}$,  carries  6 
(namely 10 components of the  metric minus 4 coordinate transformations) degrees of 
freedom in contrast to  Newton's with only one of them.  
\par
Third, in mathematical physics, especially in the search for exact solutions, 
  Birkhoff's theorem will be formulated like: up to singular  exceptions of measure
 zero, the spherically symmetric solutions of Einstein's vacuum field equation with 
 $\La$-term can be written by  inserting the value $n=2$ into the 
metric\footnote{Here, $   d\Omega^2$ denotes the metric of the standard 
$n$-sphere $S^n$, the cases with $n > 2$ are covered  here for later use.} 
  \be 
ds^2 =  - \left(1 - \frac{2m}{r^{n-1}} - \frac{\La r^2}{3} \right)dt^2 + \frac{dr^2}{1 
- 2m/r^{n-1} - \La r^2 /3} + r^2   d\Omega^2 \,   .
\ee
These mentioned singular exceptions are not only the horizons where the component 
 $g_{tt}$ of metric (1.2) vanishes, but also such solutions, where $r$ is constant, 
and therefore cannot be applied as coordinate; this can take place if the space-time
$ds^2$ represents the direct product of two 2-spaces of constant curvature, and such 
spaces can all be generated as limits of portions of space-time  metric (1.2), see
\cite{b1a}. The 
possible inclusion of several matter fields is possible, and can generally be transformed
via the Einstein equation, or other theories of gravity under discussion, to properties of 
the Ricci tensor. The usually employed formulation reads: matter fields must be spherically
symmetric. However, at least for those theories, where matter is coupled to gravity via
 the energy-momentum tensor $T_{AB}$ only, it suffices to require that  the 
energy-momentum tensor  must be spherically  symmetric. This is, in some cases, a 
really weaker assumption.  \par
And, at the end, fourth, in differential geometry, any statement of the type: 
Every member of a family of pseudo-Riemannian space-times has more isometries 
than expected from the original metric ansatz, carries the name Birkhoff-type theorem. 
 It is this type of theorems we want to develop further; including them  there will
 be some counterexamples: families of space-times where no Birkhoff-type 
theorem  is valid. These counterexamples further confirm the conjecture, that the 
Birkhoff-type theorems have their origin in the property, that the two eigenvalues
of the Ricci tensor  of two-dimensional pseudo-Riemannian spaces always coincide,
a property not having an analogy in higher dimensions. Hence, Birkhoff-type theorems
exist only for those physical situations which are  reducible to two dimensions. \par
In the present paper, we shall try to put a further impetus to developing this fourth 
aspect of the Birkhoff theorem, namely to the question: under what circumstances, the 
solutions of a set of gravitational field equations have one   more symmetry than  
should have been expected from the assumed metric ansatz? \par
Let us elucidate this  by using  Einstein's theory without matter as  example.  
Let $ds^2$ be an Einstein space, i.e. a vacuum solution of the Einstein field
 equation with  arbitrary $\La$-term,  
of  differentiability class $C^2$ having the form
 \be 
ds^2 = d\sigma^2 + r^2 d\O ^2
\ee
where $d\sigma^2$ and $ d\O ^2 $ are pseudo-Riemannian manifolds  of dimension 
$k\ge 0$ and $n\ge 0$ respectively and arbitrary signature, and  let $r \ge 0$ 
be a scalar function\footnote{Of course, essentially $r>0$, and only such 
 isolated zeroes of the function $r$  are  
allowed which are compatible with the  requested   differentiability class.} 
on $d\sigma^2$.  Furthermore, let $ d\O ^2 $ be a space of constant curvature with 
curvature scalar $R$, that means, the dimension of the  isometry group of 
$ d\O ^2 $ equals $n(n+1)/2$. In the most important application, $d\O ^2$ is the
metric on the unit sphere $S^n$, and then metric (1.3) is called to represent a 
spherically symmetric metric. If additionally $k=n=2$ and $d\sigma^2$ has 
signature $(-,+)$, this metric is called to represent a spherically symmetric space-time. 
 \par  Then the questions arise: For what values 
of $k$ and $n$ can we prove that  $D \ge 1 + n(n+1)/2$, where $D$  is the 
dimension of the local isometry group\footnote{A group $G$ is called to be a local 
isometry group of  a pseudo-Riemannian  manifold $M$, if each point of $M$
 possesses an open neighbourhood $U$ which is isometric to an open  subset 
of  a pseudo-Riemannian  manifold $V$ possessing $G$ as isometry group.} 
 of $ds^2$? That means, does the resulting  $ds^2$ possess at least one non-expected
 isometry? Does the result depend on  the signatures?  Does the result depend on the 
sign of $R$? An affirmative  answer is well-known for $k = n = 2$ and $R >0$, for 
this case  the result is called Birkhoff  theorem. This fact motivates our notation: \par
The type($k,n$)-Birkhoff theorem states the following: if metric (1.3) represents
 an Einstein space\footnote{An Einstein space is  a space  whose trace-free part of 
the Ricci tensor  vanishes.}, then the dimension of  its local isometry group is larger 
than the  dimension of the isometry group of the prescribed $ d\O ^2 $. In section 2 we 
shall outline, for which values of $k$ and $n$, a type($k,n$)-Birkhoff theorem 
 is valid. Section 3 shall show in more details the necessary formulas, section 4 
presents the coordinate-free proof of Birkhoff's   theorem for $k=2$, and the final
 section 5 shows how the metric of the generalized Schwarzschild-de Sitter black hole 
 can be deduced and gives a summary of results and some further comments.   \par
Before we continue  with answering  this type of questions  in the next sections, 
 here we present a very short overview about other papers on that theorem:
The Birkhoff  theorem for Einstein's general relativity theory has been
discussed e.g. in  \cite{b2}, \cite{b4}: here  the original 
paper:  G. D. Birkhoff,  {\it  Relativity and Modern Physics}, Harvard University Press,
 Cambridge (1923) is cited as follows: "The field outside of the spherical distribution 
of matter is static whether or not the matter  is in a static or in a variable 
state \dots Thus the Schwarzschild solution   is essentially the most general 
solution of the field equations with  spherical symmetry.", \cite{b5}: here, a 
 completely covariant proof  is given without the necessity to introduce special
 adapted coordinates; that proof shows the geometric origin of the 
Birkhoff  theorem: it rests on the property, that differently from all other dimensions 
 $k$ it holds for $k=2$: the Ricci tensor has no more than $k-1$ different 
eigenvalues, and this property has to be applied to the space perpendicular to
 the orbits of the spherical symmetry, \cite{b6}, 
\cite{b49}, \cite{b50}, \cite{b52}, \cite{b61}, \cite{b62},
 \cite{b72}, \cite{b11}, \cite{b58}, \cite{b64}, \cite{b16}, 
\cite{b17}, \cite{b20}, \cite{b25}: here a 5-dimensional exceptional case
 related to Birkhoff's theorem is covered by a non-trivial limit of space-times, 
\cite{b45}, \cite{b77},  \cite{b30}, \cite{b37}, and  \cite{b41}. \par
 The generalization of this theorem to fourth-order gravity is subject of the
following references: \cite{b57}: here it is shown  that the Birkhoff theorem is 
not valid in a fourth order theory of gravitation where $L=R^2$, and that in
this theory, the Newtonian limit is not well-behaved; only as a side-remark he 
mentions the possibility to use instead $L=R + l^2 \cdot R^2$ and comments this
to be arbitrary and being only an unwarranted complication of the theory, 
 \cite{b43}: here an example of  a fourth order theory of gravitation is presented,
where  the Birkhoff theorem is   valid, this is done by a Lagrangian, which coincides 
with $L=R$ in all those cases, where two of the eigenvalues of the Ricci tensor 
coincide, \cite{b44}, \cite{b7}, \cite{b53}, 
\cite{b78}, \cite{b56}, \cite{b9}: here it is outlined that for the  3-dimensional
 case, i.e. for $k+n=3$, the most general spherically symmetric metric cannot be 
presented in the form of metric (1.3); however, in the present  paper we restrict
 to spaces of the form (1.3) from the beginning, \cite{b10}: here the Birkhoff theorem 
for Lovelock gravity is proven, and in  \cite{b12},  a minor error of that paper is 
corrected, \cite{b14}, \cite{b22}, \cite{b26}, \cite{b27}, \cite{b28},
 \cite{b29}, \cite{b33}, \cite{b34}, \cite{b35}, \cite{b36}, 
\cite{b40}, \cite{b39}, \cite{b48}, \cite{b42}, \cite{b80},
 \cite{b46}, \cite{b66}, and  \cite{b63}. The special case of  
the conformally invariant Weyl theory has been dealt with in
 \cite{b74},  \cite{b3}, \cite{b47}, \cite{b59}, \cite{b18}, 
\cite{b21}, \cite{b23}, \cite{b70}, \cite{b31}, \cite{b32}, 
\cite{b69},  \cite{b68}, \cite{b71}, and  \cite{b75}. \par
 The  relation of the Birkhoff theorem to two-dimensional  space-times
is worked out in   \cite{b67},  \cite{b81}, \cite{b76}, 
 \cite{b73},  \cite{b8}, \cite{b51}, \cite{b54}, \cite{b55}, 
\cite{b82}, \cite{b13}, \cite{b79}, \cite{b15}, \cite{b60}, 
\cite{b19},  \cite{b65}, \cite{b24}, and  \cite{b38}. 
 For related work on black holes in Palatini gravity see e.g. \cite{b86}.
 Several variants of Birkhoff-type theorems including those in  higher dimensions and
 those including  many different types of matter fields are presented in \cite{b85}. 
The inverted Birkhoff theorem  is the subject of references  \cite{b83} and \cite{b84}.

\section{Arbitrary dimension of the spaces}
\setcounter{equation}{0} 
Let us now return to the question posed in section 1, and discuss the different cases.
 The first two ones are  trivial, we mention them only  for completeness.  \par 
First case:  $k=0$, so with eq. (1.3) we have
$ds^2 =  r^2 d\O ^2$ with a constant $r$, as  $d\sigma^2$ represents a 
one-point set only. Hence, $D=n(n+1)/2$, and so this case is not possible. \par 
Second  case: $k >0$ and $n=0$,  with eq. (1.3) we have
$ds^2 =  d\sigma^2$, and the scalar $r$ does not enter the equations. 
For $k=1$ we get $D=1 >0$, so this case is possible, it expresses the well-known 
fact that 1-dimensional Riemannian spaces are always locally homogeneous
Einstein spaces. For $k=3$ we get $D=6 >0$,  so this case is possible, too,
it expresses the fact that 3-dimensional Einstein spaces are always locally
 of constant curvature.  For all other values $k$ however, and every signature, 
 Einstein spaces without local isometries exist, so these cases are not possible. 
In sum up to now: If $k \cdot n =0$, then exactly the type($1,0$)-Birkhoff theorem 
 and the type($3,0$)-Birkhoff theorem are valid.
From now on we will  assume $k > 0$ and $n>0$. \par
 Third  case: $k=1$. For $n=1$ the essential metric form is
 $ds^2 = dx^2 + r^2(x) dy^2$ which is always an  Einstein space, but has typically 
only $D=1$, so the type($1,1$)-Birkhoff theorem is not  valid.
 For $n=2$, we again apply the fact that 3-dimensional Einstein spaces are 
 locally of constant curvature, i.e.  the type($1,2$)-Birkhoff theorem is valid.
 For $n \ge 3$, there always exist Ricci-flat spaces of the required form
 with only one Killing vector, so no Birkhoff  theorem of one of  these types  is valid.
\par
 Fourth  case: $k=2$. As is generally known, the type($2,n$)-Birkhoff theorems  
are valid for every $n \ge 1$. For a similar and in some respect more general 
approach in the context of multidimensional  gravity see \cite{b85}. \par
Fifth  case: $k \ge 3$. This case cannot be adequately dealt by such general 
considerations, so we must go deeper into the details; this we will do in the next sections. 

\section{General warped product}
\setcounter{equation}{0}  
We start with metric (1.3), representing a warped product with warping function 
$r^2$  equipped with coordinates $x^A$, where $A, \, B = 1, \dots, N$
 \be 
ds^2 = d\sigma^2 + r^2 d\O ^2 = g_{AB} dx^A dx^B \, .
\ee
With $N = k+n$ and $i,  \, j = 1, \dots, k $ we assume both $r$ and $g_{ij}$
 to depend on the $x^i$ only, and 
\be 
d\sigma^2 = g_{ij} dx^i dx^j, \qquad  r = e^\varrho, \quad r \ge 0. 
\ee
Consequently, we get for the other part of the metric
\be 
 r^2 d\O ^2 = g_{\a\b} dx^\a dx^\b \, ,
\ee
where $\a, \, \b = k+1, \dots, N $.  \par 
Now we perform a conformal transformation as follows
\be 
d\hat s^2 = e^{-2\varrho} ds^2 = h_{AB} dx^A dx^B \, .
\ee
Therefore, 
\be 
 h_{AB} = e^{-2\varrho}  g_{AB}, \qquad d\O^2 = h_{\a\b} dx^\a dx^\b 
\ee
and the $h_{\a\b}$ shall depend on the $x^\a$ only. Consequently,
\be 
d\sigma^2 = e^{2\varrho} h_{ij} dx^i dx^j, 
\ee
and
\be 
d\hat s^2 = e^{-2\varrho} d\sigma^2 + d \O^2
\ee
represents a direct product, so its Ricci tensor $P_{AB}$ has a block structure composed 
from $P_{ij}$ and $P_{\a\b}$, whereas all values $P_{i\a}$  identically vanish. Let
\be 
P = h^{AB} P_{AB} \qquad {\rm and} \qquad Q = h^{\a\b} P_{\a\b}
\ee
that means, $P$ is the curvature scalar for $d\hat s^2$ and $Q$ 
is the curvature scalar for $d\O^2$. Let $d$ be the dimension of the local isometry 
group of $d\O^2$. This implies that $d \le n(n+1)/2$ with equality taking place only
 for spaces $d\O^2$ being locally of constant curvature. \par
 Now, for a given $d\O^2$ but unspecified $r$ and $d\sigma^2$ we request $ds^2$
 to be an Einstein space. Let $D$ be the dimension of the local isometry group of 
$ds^2$. For those cases where we get $D >d$, we have the validity of a 
Birkhoff-type theorem.\footnote{Of course, $D \ge d$ follows already from the assumptions, 
so the main point is,  that for a special metric ansatz, the validity of Einstein's vacuum 
equation with $\La$-term shall imply the existence of at least one further isometry.}  \par
The case $N=0$ is trivial: for this case, we have $D=d=0$, and no Birkhoff theorem is 
valid. So we assume $N \ge 1$ in the following. Let $R_{AB}$ be the Ricci tensor of 
$ds^2$ and $R = g^{AB}R_{AB}$ the related scalar. According to our request we have 
\be 
R_{AB} = \frac{R}{N} \,  g_{AB}\, .
\ee
The case $N=1$ can now be solved: eq. (3.9) is no additional requirement, as $R=0$
 and $R_{AB}=0$ anyhow. This implies $D=1$, because every 1-dimensional 
 Riemannian space has a translational isometry, at least locally.  Further,  for 
$n=1$ we have $d=1$, and no Birkhoff  theorem holds. For $n=0$  we have $d=0$, 
and a  Birkhoff  theorem is formally valid, but it carries no more  information, than the
well-known fact, that every one-dimensonal Riemannian space has an isometry.  
The case $N=2$ is like-wise trivial: every two-dimensional pseudo-Riemannian space
is an Einstein space, so requesting it is an empty requirement, and cannot increase the 
dimension of the isometry group. So, for $N=2$  no Birkhoff  theorem holds. \par
Let us assume $N \ge 3$ in  the following, this implies $R$ to be constant, see eq. (3.9). 
   The conformal transformation eqs. (3.4)/(3.5) has the 
following consequence  for the related Ricci tensors: 
\be
P_{AB}= R_{AB}+(N-2)\left(\varrho_{;AB}+\varrho_{;A}\varrho_{;B}\right)
+ g_{AB}\left(\Box \varrho - (N-2)\varrho^{;C}\varrho_{;C} \right) \, ,
\ee
where indices are moved and covariant derivatives $f_{;AB}$ are calculated  with the 
metric $g_{AB}$, and the D'Alembertian $\Box$  is defined by   $\Box f=  g^{AB}f_{;AB}$ 
for any scalar $f$.  Of course, $f_{;A}$ is identical to the partial derivative $f_{,A}$.  \par
Transvecting eq. (3.10) with $g^{AB}$ we get 
\be
P \cdot e^{-2\varrho} = R + (N-1) \left(2 \Box \varrho - 
(N-2)\varrho^{;C}\varrho_{;C} \right) \, .
\ee
To calculate $\varrho_{;AB}$, we need the components of the Christoffel affinity.
 To this end we rewrite eq. (3.1) as follows:
\be 
ds^2 = g_{ij} dx^i dx^j + r^2 h_{\a\b} dx^\a dx^\b 
\ee
with $g_{ij}$  and $r$ depending on $x^i$ only, whereas the  $h_{\a\b}$ depend 
on the $ x^\a$ only. It is clear from the above, that we have also
$$
r^2 h_{\a\b}= e^{2\varrho} h_{\a\b} = g_{\a\b} \, .
$$
The components $\Gamma^i_{jl}$ of the Christoffel affinity represent both the complete
 Christoffel affinity for the space $d\sigma^2$  as well as those components with indices 
all $\le k$ of the  Christoffel affinity for the space $ds^2$. In our case, both interpretations 
lead to the  same values, so we need not distinguish the notation here. \par
 The same takes  place with the components $\Gamma^\a_{\b\gamma}$ 
of the Christoffel affinity: in the following three spaces their value is always the same:
for $  d \O^2$, for $ r^2 d \O^2$, and for the components with indices all $> k$ 
of the  Christoffel affinity for the space $ds^2$. \par 
The only non-trivial influence of  an allowed non-constancy of the warping factor 
$r^2=e^{2\ro}$ is via the following components of the Christoffel affinity for the space $ds^2$:
\be 
\Gamma^\a_{\b i} = \delta^\a_\b \, \varrho_{;i} \, \qquad {\rm and} \qquad
\Gamma^i_{\a\b} = - g_{\a\b} \varrho^{;i}\, . 
\ee
Now we are ready to calculate the needed components of $\varrho_{;AB}$:
all the mixed components $\varrho_{;\a i}$ vanish, the components 
$\varrho_{;ij}$ can be calculated if they were simply within $d\sigma^2$, and the only
non-trivial part is 
\be 
\varrho_{;\a \b} = g_{\a\b}  \varrho^{;i} \varrho_{;i}  \, . 
\ee
Denoting the D'Alembertian within $d\sigma^2$ by $\Delta$, i.e.
  $\Delta \varrho = g^{ij}\varrho_{;ij} $ we get
\be 
\Box \varrho = \Delta \varrho + n \cdot  \varrho^{;i} \varrho_{;i}  \, . 
\ee
By construction, see eq. (3.8), $Q$ depends on the $x^\a$ only, and $P-Q$, 
representing the curvature scalar for $e^{-2\ro}d\sigma^2$, depends on the 
$x^i$ only. Now we are ready to evaluate eq. (3.10) in more details: Inspection of 
the mixed components  implies that $R_{\a i} = 0$ identically. So, we may split 
eq. (3.10) in the $\a\b$-block and the $ij$-block. So we get
\be 
P_{\a\b} = R_{\a\b} + g_{\a\b} \Box \ro
\ee
and
\be 
P_{ij} = R_{ij} + (N-2)\left(\varrho_{;ij}+\varrho_{;i}\varrho_{;j}\right)
+ g_{ij}\left(\Box \varrho - (N-2)\varrho^{;l}\varrho_{;l} \right) \, .
\ee
Inserting  eq. (3.9) into these two equations  we get 
\be 
P_{\a\b} =  g_{\a\b}  \left(   \Box \ro  + \frac{R}{N}   \right)
\ee
and
\be 
P_{ij} =  (N-2)\left(\varrho_{;ij}+\varrho_{;i}\varrho_{;j}\right)
+ g_{ij}\left(\Box \varrho + \frac{R}{N} - (N-2)\varrho^{;l}\varrho_{;l} \right) \, .
\ee
Transvecting eq. (3.18) with $h^{\a\b}$ we get with eq. (3.8)
\be 
Q =  n e^{2\ro}   \left(   \Box \ro  + \frac{R}{N}   \right) \, . 
\ee
Transvecting eq. (3.19) with $h^{ij}$ we get with eq. (3.8)
\be 
P - Q = e^{2\ro} \left((2N-n-2) \Box \ro  + \frac{R}{N}(N-n)
- (N-1)(N-2) \varrho^{;i}\varrho_{;i}   \right) \, . 
\ee
Cross-checking of eqs. (3.20)/(3.21) with eq. (3.11) shows that these 3 equations 
are compatible. \par
By construction, the following terms depend on the $x^i$ only: $\ro$, 
$\varrho^{;i}\varrho_{;i}$, $\Delta \ro$, $\Box \ro$, $P-Q$, $g_{ij}$, 
$h_{ij}$, $P_{ij}$, and $R_{ij}$. Likewise by construction, the following terms 
depend on the $x^\a$ only: $Q$, $P_{\a\b}$, and $h_{\a\b}$.  \par
We get the result: the l.h.s. of eq. (3.20) depends on the $x^\a$ only, 
and its r.h.s. depends on the $x^i$ only. Consequently, $Q$ is a constant. 

\section{Coordinate-free proof of Birkhoff's theorem}
\setcounter{equation}{0}  
Now we restrict to the main case, $k=2$, i.e. $N=n+2$. The relevant equations 
from section 3 then lead to the following simplifications: From eq.  (3.9) we get
\be 
R_{AB} = \frac{R}{n+2} \,  g_{AB}  \qquad R= {\rm const. }
\ee
Eq. (3.18) now reads
\be 
P_{\a\b} =  g_{\a\b}  \left(   \Box \ro  + \frac{R}{n+2}   \right)\, ,
\ee
from eq. (3.19) 
we 
get
\be 
P_{ij} =  n\left(\varrho_{;ij}+\varrho_{;i}\varrho_{;j}\right)
+ g_{ij}\left(\Box \varrho + \frac{R}{n+2} - n\varrho^{;l}\varrho_{;l} \right) \, .
\ee
From eq. (3.20) we get 
\be 
Q =  n e^{2\ro}   \left(   \Box \ro  + \frac{R}{n+2}   \right) = {\rm const. }
\ee
From eq. (3.21)  we get 
\be 
P - Q = e^{2\ro} \left((n+2) \Box \ro  + \frac{2R}{n+2}
- n(n+1) \varrho^{;i}\varrho_{;i}   \right) \, . 
\ee
At this place it proves useful to re-insert $r=e^\ro$ instead of $\ro$ into 
the equations: $\ro = \ln r$, and similarly we get with eq.  (3.15)
$$
\varrho^{;i}\varrho_{;i} = \frac{1}{r^2} \cdot r^{;i}r_{;i}, \qquad 
\Box \ro = \frac{\Delta r}{r} + \frac{n-1}{r^2}\cdot r^{;i}r_{;i}\, .
$$
then eqs. (4.2) - (4.5) read
\be 
P_{\a\b} =  g_{\a\b}  \left(\frac{\Delta r}{r} +
  \frac{n-1}{r^2} \cdot r^{;i}r_{;i}   + \frac{R}{n+2}   \right)\, ,
\ee
\be 
P_{ij} =  \frac{n}{r} \cdot r_{;ij} + g_{ij}\left(\frac{\Delta r}{r} -
  \frac{1}{r^2} \cdot r^{;i}r_{;i}   + \frac{R}{n+2}  \right) \, .
\ee
\be 
Q =  n  \left(  r  \Delta r  + (n-1) \cdot r^{;i}r_{;i}
 + \frac{r^2 \cdot R}{n+2}   \right) = {\rm const. }
\ee
\be 
P - Q = (n+2) r \Delta r   + \frac{2r^2 R}{n+2} - 2 r^{;i}r_{;i}    \, . 
\ee
What can we directly see here is the following: if we insert 
 $ g_{\a\b} = r^2  \cdot  h_{\a\b}$  into eqs. (4.6) and (4.8) we get 
\be 
P_{\a\b} =  \frac{Q}{n} \cdot h_{\a\b}, \qquad  Q = {\rm const. }
\ee
That means, $d\Omega^2$ is an Einstein space with constant curvature
scalar. It is essential to point out that we have not assumed   
$d\Omega^2$ to be an Einstein space, or even a space of constant curvature,
but moreover, it follows from the other assumptions; of course, the constancy 
 of $Q$ is a non-trivial extra property for $n=2$ only.  \par
The antisymmetric Levi-Civita pseudo-tensor $\e_{ij}$ in $d\sigma^2$ is 
completely defined by $\e_{12} = \sqrt{\vert  \det g_{ij} \vert }$. It is
covariantly constant.  We
 now define the pseudo-vector $\xi^i$ via
\be 
\xi_i = \e_{ij} \ r^{;j} \, . 
\ee
Here is  the most relevant point of the deduction: In two-dimensional 
 spaces, the two eigenvalues of the  Ricci tensor coincide.\footnote{And, by the way, 
 just equal the Gaussian curvature of the surface.} Therefore, $P_{ij}$ is 
proportional to $g_{ij}$, and with eq. (4.7) we see that this also takes place for
$r_{;ij}$. So we insert $r_{;ij} = c \cdot g_{ij}$ with a scalar $c$ into eq. (4.11)
and get finally $\xi_{i;j} + \xi_{j;i} =0$. Hence, $\xi_i$ is a Killing vector.\footnote{Of
course, formally it is a pseudo-vector only, but replacement of  $\xi_i$ by
$-\xi_i$ does not alter the Killing equation, so we may keep the word  Killing vector.}

\section{Discussion}
\setcounter{equation}{0}  
Now we can summarize the results in the following \par
{\bf Generalized Birkhoff Theorem}: Let the warped product
\be 
ds^2 = d\sigma^2 + r^2 d\Omega^2
\ee
be an Einstein space, where $d\sigma^2$ is two-dimensional with coordinates
 $x^i$, and $d\Omega^2$ is $n$-dimensional with $n \ge 1$. The warping factor $r^2$
depends on the $x^i$ only. Then it holds: $\xi_i = \e_{ij} \ r^{;j}$ represents a
hypersurface-orthogonal Killing vector for $ds^2$. Hence, the dimension of the isometry 
group of $ds^2$ is larger than  the dimension of the isometry group of $d\Omega^2$. \par
{\bf  Proof}: That  $\xi_i $ is a Killing vector in  $d\sigma^2$ was already deduced 
earlier, and in 2 dimensions, every vector is hypersurface-orthogonal anyhow. 
That both properties are maintained if $\xi_i $ is  lifted to $ds^2$ becomes clear from
the construction. If $r_{;i}$ vanishes on a hypersurface only, then so does $\xi_{i}$,
 but this does not prevent $\xi_{i}$ to induce an isometry, as it remains non-zero
 in a dense subset of the manifold. If $r_{;i}$ is a non-vanishing light-like vector in 
a whole region, then $d\sigma^2$ is flat, see the first paper in  \cite{b67}, sct. V A,
so 3 Killing vectors appear.  It still remains to look for the case, that $r_{;i}=0$
 in a whole region. Then with eq. (4.7) we get $P_{ij} = g_{ij} R/(n+2)$ with constant $R$,
so $d\sigma^2$ must be a space of constant curvature, it possesses 3 independent 
Killing vectors. q.e.d.   \par
{\bf  Summary of results}: For the metric (1.3), reading $ds^2 = d\sigma^2 + r^2 d\O ^2$,
where $d\sigma^2$ and $ d\O ^2 $ are pseudo-Riemannian manifolds  of dimension 
$k\ge 0$ and $n\ge 0$ respectively and arbitrary signature, and  where $r$ lives on
$d\sigma^2$, we prescribe $ d\O ^2 $ with a $d$-dimensional isometry group.
 Concerning $r$ and $d\sigma^2$ we only require that $ds^2$ is an Einstein space. 
Let $D$ be the dimension of the isometry group of $ds^2$. Then the
 type$(k,n)$-Birkhoff theorem states that $D \ge d+1$.
\vspace*{1cm}
$$
(1,0), \, (3,0), \, (1,2), \, (2,n), \, n \ge 1 
$$
\vspace*{1cm}
Table 1: Values of $(k,n)$, where the type$(k,n)$-Birkhoff theorem is valid
\vspace*{1cm}
$$
(0,n), n \ge 0, \, (2,0), \, (k,0), \, k \ge 4, \, (1,1), \,  (1,n), n \ge 3, \, (k,n), \, k \ge 3
$$
\vspace*{1cm}
Table 2: Values of $(k,n)$, where the type$(k,n)$-Birkhoff theorem is {\it not} valid
\vspace*{1cm}

\par
{\bf   Comments}: 1. As a byproduct we have shown that under the conditions 
of the theorem, $ d\Omega^2$  turns out to  be an Einstein space with constant
 curvature scalar, a property, which is presupposed in many other presentations.  \par
2. No similar consideration is possible for dimensions $k \ge 3$, as for spaces of 
dimension $\ge 3$ the Ricci tensor may have $k$ different eigenvalues. Therefore, 
no generalization of the Birkhoff theorem into this direction is to be expected. \par 
3. Differently from other proofs, we did not introduce any coordinates. 
Besides aesthetic reasons, this approach has the great advantage, that no special
 care is needed to deal with the horizon. Let us make this point more detailed: 
 In regions, where $r_{;i}$ is different from zero, one could be tempted to use $r$
as one  coordinate, and to define the other coordinate, denoted by $t$, by the 
condition: the $t$-lines shall always be perpendicular to the $r$-lines. But then 
immediately it becomes clear, that for light-like values $r_{;i}$, the coordinate $t$
is not well-defined.\footnote{In more details: That the $t$-lines are perpendicular
to the $r$-lines can be expressed by the condition $t^{;i}r_{;i}=0$, but if
$r_{;i}$ is a non-vanishing light-like vector, then  $t^{;i}$ must be 
parallel to $r^{;i}$, so the coordinates $r$, $t$ fail to be independent ones.} \par
4. To check the applicability of the presented formulas\footnote{and also to increase 
the confidence in their correctness}, let us  try to deduce the higher-dimensional 
Schwarzschild-de Sitter solution eq. (1.2) using the presented approach. 
Now it is indeed worthwhile to introduce the coordinates $r$ and $t$ for $ds^2$
as described in the previous comment, knowing that we now do not cover
those points of the manifold, where $r_{;i}$ changes from space-like to
time-like.\footnote{Of course, these are the same  points of the manifold, 
where $\xi_{i}$ changes from time-like to space-like, i.e., the points of the horizon.} 
By construction, $g_{12}=0$, and the other components $g_{ij}$ depend on $r$
only. Restricting now to time-like $x^1 = t$ and space-like $x^2 = r$ only, we can 
now write
$$
d\sigma^2 = - A(r) dt^2 + \frac{dr^2}{B(r)}
$$
with positive functions  $A(r)$ and $B(r)$. Now we skip the standard argument
that shows that putting $A(r) = B(r)$ does not restrict generality in this context. 
So we use eq. (5.1) with
\be 
d\sigma^2 = - A(r) dt^2 + \frac{dr^2}{A(r)}
\ee
and $d \Omega^2$ being the metric of the standard sphere $S^n$, and 
we restrict to the case $n \ge 2$.
 With a dash denoting the  derivative with respect to $r$ we get
 $\Delta r =  A'(r)$ and $r_{;i}r^{;i} = A(r)$. With eq. (4.8) we get then
$$
\frac{Q}{n} = r A'(r) + (n-1) A(r) + \frac{r^2 R}{n+2}\, .
$$
To solve this equation it proves useful to define the function $F(r) = r^{n-1} A(r)$. 
We simply get 
$$
F'(r) = \frac{Q}{n} \cdot r^{n-2} - \frac{r^n R}{n+2}
$$
which can be integrated to 
$$
F(r) = c_1 + Q_1 r^{n-1} - R_1 r^{n+1}
$$
with constants $c_1$, $Q_1$ and $R_1$, where $Q$ and $Q_1$ have the same 
sign, and $R$ and $R_1$ have the same sign. Finally we get
 \be 
A(r) = Q_1 + \frac{c_1}{r^{n-1}} - R_1 r^2 \, . 
\ee
Comparison with eq. (1.2) clearly shows the physical interpretation of the 
three constants in metric (5.1) with (5.2) and (5.3).

\end{document}